\documentstyle[12pt,epsf]{article}
\begin{document}
\baselineskip 0.7cm

\newcommand{\gsim}{ \mathop{}_{\textstyle \sim}^{\textstyle >} }
\newcommand{\lsim}{ \mathop{}_{\textstyle \sim}^{\textstyle <} }
\newcommand{\vev}[1]{ \left\langle {#1} \right\rangle }
\newcommand{\lsp}{ \left ( }
\newcommand{\rsp}{ \right ) }
\newcommand{\lmp}{ \left \{ }
\newcommand{\rmp}{ \right \} }
\newcommand{\llp}{ \left [ }
\newcommand{\rlp}{ \right ] }
\newcommand{\labs}{ \left | }
\newcommand{\rabs}{ \right | }
\newcommand{\KEV}{ {\rm keV} }
\newcommand{\MEV}{ {\rm MeV} }
\newcommand{\GEV}{ {\rm GeV} }
\newcommand{\TEV}{ {\rm TeV} }
\newcommand{\mgut}{M_{GUT}}
\newcommand{\mint}{M_{I}}
\newcommand{\mgra}{M_{3/2}}
\newcommand{\mll}{m_{\tilde{l}L}^{2}}
\newcommand{\mdr}{m_{\tilde{d}R}^{2}}
\newcommand{\mllXX}[1]{m_{\tilde{l}L , {#1}}^{2}}
\newcommand{\mdrXX}[1]{m_{\tilde{d}R , {#1}}^{2}}
\newcommand{\mgy}{m_{G1}}
\newcommand{\mgl}{m_{G2}}
\newcommand{\mgc}{m_{G3}}
\newcommand{\nuR}{\nu_{R}}
\newcommand{\slL}{\tilde{l}_{L}}
\newcommand{\slLi}{\tilde{l}_{Li}}
\newcommand{\sdR}{\tilde{d}_{R}}
\newcommand{\sdRi}{\tilde{d}_{Ri}}
\newcommand{\e}{{\rm e}}
\newcommand{\Br}{{\rm Br}}

\begin{titlepage}

\begin{flushright}
KEK-TH-548,UT-799
\\
November, 1997
\end{flushright}

\vskip 0.35cm
\begin{center}
{\large \bf  Atmospheric Neutrino Oscillation \\
            and  \\
             Large Lepton-Flavour Violation\\
            in the SUSY SU(5) GUT}
\vskip 1.2cm
J.~Hisano$^{a)}$, Daisuke Nomura$^{b)}$, and T.~Yanagida$^{b)}$
\vskip 0.4cm

a) {\it Theory Group, KEK, Tsukuba, Ibaraki 305, Japan}
\\
b) {\it Department of Physics, University of Tokyo, Tokyo 113, Japan}
\vskip 1.5cm

\abstract{
The atmospheric neutrino anomaly reported by the super-Kamiokande
collaboration suggests existence of a large flavour violating Yukawa coupling
 in the lepton sector. We discuss lepton-flavour violation at low energies in 
the framework of the supersymmetric $SU(5)_{GUT}$ model with right-handed
 neutrinos. We find that for a wide range of parameter space suggested 
 from the atmospheric neutrino anomaly, if the tau-neutrino Yukawa coupling 
is as large as that of top quark, lower bounds of the 
branching ratios of $\mu^+ \rightarrow \e^+ \gamma$, $\tau^\pm \rightarrow 
\mu^\pm \gamma$, and the $\mu-\e$ conversion rate  on  $^{48}_{22}{\rm Ti}$ 
become around 
$10^{-14}$, $10^{-9}$, and $10^{-16}$, respectively. These reaction rates
 may be in the region accessible
by near future experiments.

}

\end{center}
\end{titlepage}

Small dimensional parameters in physics were often
important indications of new, more fundamental physics in the past.
A neutrino mass is considered as such an example in the present particle
physics. The see-saw mechanism \cite{seesaw} is the most attractive theory for 
light massive neutrinos in which the smallness of their masses is naturally 
explained by large Majorana masses of their chiral partners (i.e. 
right-handed neutrinos).

At present there are two experimental hints of nonvanishing neutrino masses:
one is the well-known solar neutrino deficit \cite{solar_neutrino} and the 
other is the atmospheric neutrino anomaly \cite{atm_neutrino}. A recent report on
the atmospheric neutrino from the super-Kamiokande collaboration 
\cite{superkamiokande} has presented 
convincing evidence that the atmospheric neutrino anomaly is indeed due to 
neutrino oscillation. From the zenith-angle dependence of the ratio of 
$\nu_e$ and $\nu_{\mu}$ fluxes the super-Kamiokande collaboration 
\cite{superkamiokande} has suggested a neutrino mass difference as 
   \begin{equation}
     \delta m^2_{\nu_{\mu} \nu_X} \simeq (10^{-3} - 10^{-2}) {\rm eV^2} .
   \end{equation}

Provided mass hierarchy $m_{\nu_{\tau}}> m_{\nu_{\mu}}> m_{\nu_e}$
 it is natural to consider $\nu_X = \nu_{\tau}$ while the solar neutrino 
deficit is explained by the matter oscillation between $\nu_{\mu}$ and
$\nu_e$  \cite{MSW}. Thus, the atmospheric neutrino anomaly implies a 
nonvanishing neutrino mass,
   \begin{equation}
      m_{\nu_{\tau}} \simeq (0.03-0.1)  {\rm eV} ,
   \end{equation}
which suggests the presence of right-handed neutrinos at the scale 
$\sim$ $(10^{14}-10^{15})$ GeV. This may support strongly 
supersymmetric (SUSY) grand unified theories (GUT's), since the mass of the 
right-handed neutrino is
very close to the GUT scale $\simeq 10^{16} {\rm GeV}.$

The other important feature of the super-Kamiokande observation is that a large
mixing
   \begin{equation}
       \sin^2 2\theta_{\nu_{\mu} \nu_{\tau}} \simeq 1 
        \label{angle}
   \end{equation}
is required to account for the atmospheric neutrino anomaly. This result 
already indicates a large lepton-flavour violation (LFV) at the GUT scale. 
In this letter we point out that Yukawa interactions for the right-handed 
neutrinos in the SUSY $SU(5)_{GUT}$ model induce large LFV at low energies.
The present model is an $SU(5)_{GUT}$-extension of the previous one 
\cite{HMTYY,HMTY}, 
%\footnote{
%
%The LFV has been also discussed in Ref.\cite{BH} which has not, however,
%considered the dominant contribution.}
%
which yields a significant enhancement in the LFV processes. We find that 
lower bounds for the branching ratios of $\mu \rightarrow e \gamma$ and 
$\tau \rightarrow \mu \gamma$, and the $\mu-\e$ conversion rate on  
$^{48}_{22}{\rm Ti}$ reach to $10^{-14}$, $10^{-9}$, and $10^{-16}$, respectively,
for a reasonable parameter region suggested from the atmospheric neutrino 
anomaly. This may be accessible by near future experiments.

We should note here that the LFV is also predicted in the minimum SUSY 
$SU(5)_{GUT}$ model without right-handed neutrinos \cite{BH}. However,
the reaction rates of the LFV processes are much smaller than predictions
in the present model as explained later (see also Ref. \cite{HMTYSU5}).

We introduce singlet chiral multiplets ${\bf 1}$ in the SUSY $SU(5)_{GUT}$ 
model in addition to the ordinary quark and lepton chiral multiplets
${\bf \bar{5}}$ and ${\bf 10}$. The singlets correspond to chiral multiplets
 for the right-handed neutrinos. Yukawa couplings of Higgs multiplets
$H({\bf 5})$ and $\overline{H}({\bf \bar{5}})$ are given by
\begin{equation}
      W_{SU(5)} =   \frac14 f_u^{ij}{\bf 10}_i {\bf 10}_j H 
          + \sqrt{2} f_d^{ij} {\bf 10}_i {\bf \bar{5}}_j  \overline{H} 
          + f_{\nu}^{ij} {\bf 1}_i  {\bf \bar{5}}_j  H ,
\label{su5}
\end{equation}
where $i,j(=1-3)$ are generation indices. 
Without loss of generality we can always take 
a basis where $f_u^{ij}$ and $f_{\nu}^{ij}$ are diagonal. In this basis 
the Yukawa couplings can be written as 
  \begin{eqnarray}
      f_u^{ij} &=& f_{u_i} {\e}^{i\phi_{u_i}} \delta^{ij}, \nonumber\\
      f_d^{ij} &=& V^{\star}_{ik} f_{d_k} U^{\dagger}_{ kj}, \nonumber\\
      f_{\nu}^{ij} &=& f_{\nu_i} {\e}^{i\phi_{\nu_i}} \delta^{ij},
\label{basis}
  \end{eqnarray}
where $V$ is the Cabbibo-Kobayashi-Maskawa (CKM) matrix of 
the quark sector, and $U$ is that of the lepton sector. The additional 
phases $\phi_{u_i}$ and $\phi_{\nu_i}$ satisfy 
$\phi_{u_1}+\phi_{u_2}+\phi_{u_3}=0$ and  $\phi_{\nu_1}+\phi_{\nu_2}
+\phi_{\nu_3}=0$.\footnote{
The Yukawa coupling constants $f_u$, $f_d$, and $f_\nu$
have $2\times(6+9+9)-3\times 9=21$ degrees of freedom up to redefinition of
 fields. These correspond to $(3\times 3+4\times2 +2\times 2)$, in which 
the first number denotes the number of the eigenvalues of the Yukawa 
coupling matrices, the second the number of parameters in CKM matrices for
quark and lepton sectors, and the third represents the additional phases 
$\phi_{u_i}$ and $\phi_{\nu_i}$.}
In the following discussion these phase parameters are irrelevant  
to our results. From Eq.~(\ref{basis}) we see that LFV interactions for 
left-handed leptons ($l_L$) in ${\bf \bar{5}}$  and right-handed leptons 
($l_R$) in ${\bf 10}$ originate from
off-diagonal elements of the matrices $U$ and $V$, respectively.

We assume that Majorana masses for ${\bf 1}_i$'s are given by the 
following nonrenormalizable interactions;
  \begin{equation}
      W =  \frac{c_{ij}}{M_G} {\bf 1}_i {\bf 1}_j \langle \Sigma \rangle^2 ,
  \label{m_rn_mass}
  \end{equation}
where $\Sigma({\bf 24})$ is the adjoint Higgs multiplet causing the breaking 
$SU(5)_{GUT}
\rightarrow SU(3) \times SU(2) \times U(1)$ and $M_G$ is the gravitational
scale $M_G \simeq 2.4 \times 10^{18} {\rm GeV}$. For simplicity we take 
$c_{ij}=c \delta_{ij}$ in the present analysis, since our 
conclusion depends very weakly on the structure of the matrix $c_{ij}$,
 as far as there is no fine tuning between the matrix $c_{ij}$ and the Yukawa 
couplings $f_{\nu_i}$.\footnote{
When there is the following relation between the Yukawa couplings and the 
Majorana masses of the right-handed neutrinos,
\[
c_{33}:c_{32}:c_{22} \simeq f_{\nu_3}^2:f_{\nu_2}f_{\nu_3}:f_{\nu_2}^2,
\]
$\sin^2 2\theta_{\nu_{\mu} \nu_{\tau}}$ can be close to one even if 
$(3,2)$ element of $U$ is very small. However, we consider that such a
 relation between independent couplings and masses is unnatural 
without some family symmetry \cite{YF}.
}
In this case the large neutrino mixing angle $\sin\theta_{\nu_\mu\nu_\tau} 
\simeq 1/\sqrt{2} $ in Eq.~(\ref{angle})
corresponds to a large $(3,2)$ element of the matrix $U$. 
Assuming $c \simeq O(1)$ we restrict our discussion to the case of the 
Majorana masses for right-handed neutrinos being
   \begin{equation}
      M_N \simeq (10^{13} - 10^{15}) {\rm GeV} .
        \label{mn_scale}
   \end{equation}
Notice that the neutrino mass $m_{\nu_{\tau}} \simeq 
(0.03-0.1) {\rm eV}$ requires the considerably large Yukawa
coupling constant $f_{\nu_3} \simeq (0.1-2.5)$ with $M_N$ given by
Eq.~(\ref{mn_scale}). 

Let us introduce soft SUSY-breaking parameters in this model. In terms of
the $SU(5)_{GUT}$ multiplets, the soft SUSY-breaking parameters 
for squarks, sleptons, and Higgses are given by \footnote{
We neglect the Yukawa coupling $\lambda H \Sigma
\bar{H}$ and soft SUSY-breaking parameters associated with it, 
for simplicity. The detailed analysis including such terms
will be given in Ref. \cite{HNT}.}
\begin{eqnarray}
-{\cal L}_{\rm SUSY~breaking} 
&=&
(m_{\bf 10}^2)^j_i  \widetilde{\bf 10}^{i\dagger} \widetilde{\bf 10}_j
+(m_{\bf \bar{5}}^2)^j_i  \widetilde{\bf \bar{5}}^{i\dagger}     \widetilde{\bf \bar{5}}_{j} 
+(m_{\bf 1}^2)^j_i \widetilde{\bf 1}^{i\dagger} \widetilde{\bf 1}_{j} 
+m_{h}^2 h^\dagger h +  m_{\bar{h}}^2 \bar{h}^\dagger \bar{h} 
\nonumber\\
&&
+\left\{
\frac14  A_u^{ij} \widetilde{\bf 10}_i \widetilde{\bf 10}_j h
+ \sqrt{2}A_d^{ij} \widetilde{\bf 10}_i \widetilde{\bf \bar{5}}_{j} \bar{h}
+ A_{\nu}^{ij} \widetilde{\bf 1}_i \widetilde{\bf \bar{5}}_{j} h + {\rm h.c.}
\right\},
\end{eqnarray} 
where $\widetilde{\bf 10}_i$, $\widetilde{\bf \bar{5}}_i$, and 
$\widetilde{\bf 1}_i$ are scalar components of the ${\bf 10}_i$, 
${\bf \bar{5}}_i$, and ${\bf 1}_i$ chiral multiplets, respectively, and 
$h$ and $\bar{h}$ are Higgs bosons.
In the minimal supergravity these parameters are given at the 
gravitational scale by
\begin{eqnarray}
&(m_{\bf 10}^2)^j_i=(m_{\bf \bar{5}}^2)^j_i=(m_{\bf 1}^2)^j_i=\delta^i_j m_0^2,&
\nonumber\\
&m_{h}^2 = m_{\bar{h}}^2= m_0^2,&
\nonumber\\
& A_u^{ij}=f_u^{ij}a_0, \,  A_d^{ij}=f_d^{ij}a_0, \, 
 A_{\nu}^{ij}=f_{\nu}^{ij}a_0. &
\label{InitCondAtGravScale}
\end{eqnarray}
We assume the above relations in the minimal supergravity in the present 
analysis. If it is not this case, the soft SUSY-breaking parameters at the
gravitational scale themselves induce extra contributions to the LFV.
Thus, we consider that our results give theoretical lower bounds on the LFV
as far as there is no accidental cancellation.

The large Yukawa coupling constants $f_{u_3}$ and  $f_{\nu_3}$ reduce 
$(m_{\bf 10}^2)^3_3$ and $(m_{\bf \bar{5}}^2)^3_3$ significantly through 
radiative corrections. To make our point clear we neglect the Yukawa 
coupling constants except for $f_{u_3}$ and $f_{\nu_3}$ in the following 
discussion. We include, however, their effects in the numerical calculations. 
We define magnitudes of the reductions as 
\begin{eqnarray}
R_{\bf 10}         &\equiv&(m_{\bf 10}^2)^1_1     -(m_{\bf 10}^2)^3_3, 
\nonumber\\
R_{\bf \bar{5}}    &\equiv&(m_{\bf \bar{5}}^2)^1_1-(m_{\bf \bar{5}}^2)^3_3. 
\end{eqnarray}
Renormalization group equations (RGE's) for them at the one-loop level are 
given by
\begin{eqnarray}
\mu\frac{dR_{\bf 10}}{d \mu} &=& -\frac1{(4\pi)^2} 6 f_{u_3}^2 
(2 (m_{\bf 10}^2)^3_3 + \frac{1}{2}m_{h}^2), \nonumber\\
\mu\frac{dR_{\bf \bar{5}}}{d \mu} &=& -\frac1{(4\pi)^2} 2 f_{\nu_3}^2 
( (m_{\bf 1}^2)^3_3 +  (m_{\bf \bar{5}}^2)^3_3 + m_{h}^2),
\label{RGE}
\end{eqnarray}
where $\mu$ is the renormalization point. Then, at the GUT scale $\mu \simeq
10^{16} \GEV $
the squark and slepton mass matrices become
\begin{eqnarray}
(m_{\bf 10}^2)=\left(
\begin{array}{ccc} 
m_{\bf 10}^2&&\\  
&m_{\bf 10}^2&\\  
&&m_{\bf 10}^2 -R_{\bf 10}
\end{array}
\right)
,&&
(m_{\bf \bar{5}}^2)=\left(
\begin{array}{ccc} 
m_{\bf \bar{5}}^2&&\\  
&m_{\bf \bar{5}}^2&\\  
&&m_{\bf \bar{5}}^2 -R_{\bf \bar{5}}
\end{array}\right),
\nonumber\\
\end{eqnarray}
where $m_{\bf 10}^2\equiv (m_{\bf 10}^2)^1_1=(m_{\bf 10}^2)^2_2$
and $m_{\bf \bar{5}}^2\equiv (m_{\bf \bar{5}}^2)^1_1=(m_{\bf \bar{5}}^2)^2_2$.

The Yukawa coupling $f_d$ in Eq.~(\ref{basis}) splits into two parts below 
the GUT scale: one is $f_l$ for leptons and the other $f_d$ for
down-type quarks. They satisfy a relation $f_l=f_d$ at the GUT scale. 
Below the GUT scale we take a basis where the Yukawa coupling matrix 
$f_l$ for leptons is diagonalized as
\begin{eqnarray}
W_{{\rm MSSM}+\nu_R} &=& 
  f_{u_i}  Q_i\overline{U}_i H_2
+ V^{\star}_{ij} f_{d_j}  Q_i \overline{D}_j{H}_1
+ f_{l_i} \overline{E}_i L_i{H}_1,
+ f_{\nu_i} U^{ij} \overline{N}_i L_j {H}_2,
\nonumber\\
\label{MSSM}
\end{eqnarray}
where $Q$, $\overline{U}$ and  $\overline{D}$ are left-handed quark and 
right-handed quark chiral multiplets,
$L$, $\overline{E}$, and $\overline{N}$ are those for left-handed leptons, 
right-handed leptons, and right-handed neutrinos, and  $H_1$ and 
${H}_2$ are those for doublet Higgses. These quarks 
and leptons are embedded in ${\bf 10}_i$,  ${\bf \bar{5}}_i$, and ${\bf 1}_i$ as 
\begin{eqnarray}
{\bf 10}_i      &=& \left\{Q_i,  \e^{-i\phi_{ui}}\overline{U}_i, V^{ij} \overline{E}_j \right\} ,
\nonumber\\
{\bf \bar{5}}_i &=& \left\{U^{ij} \overline{D}_j,  U^{ij} L_j \right\}, 
\nonumber\\
{\bf 1}_i       &=& \left\{\e^{-i\phi_{\nu i}}\overline{N}_i \right\} .
\end{eqnarray}
Therefore, the mass matrix for the left-handed  sleptons $(m_{\tilde{l}_L}^2)$
and that for the right-handed sleptons $(m_{\tilde{l}_R}^2)$ 
are given by\footnote
{In the SUSY $SU(5)_{GUT}$ model with the right-handed neutrinos 
the right-handed down-type squarks also have flavour violating masses, 
and they can contribute to $b\rightarrow s\gamma$ through gluino diagrams. 
However, the flavour violating masses for the right-handed down-type squarks 
are the same as those 
for the left-handed sleptons at the GUT scale, which is constrained from 
$\Br(\tau\rightarrow \mu\gamma)<4.2\times 10^{-6}$ \cite{PDG}.
Then, it is difficult to generate a significant contribution to 
$b\rightarrow s\gamma$ unless we choose an unnatural parameter region.
}
\begin{eqnarray}
(m_{\tilde{l}_L}^2)^i_j &=&  U^{\dagger}_{jk}(m^2_{\bf \bar{5}})^k_l  U^{li}
\nonumber\\
                        &=&   m_{\bf \bar{5}}^2 \delta^i_j 
                           - R_{\bf \bar{5}} U^{\dagger}_{j3} U^{3i},
\\
(m_{\tilde{l}_R}^2)^i_j &=&  V^{\dagger}_{jk}(m^2_{\bf 10})^k_l  V^{li}
\nonumber\\
                           &=& m_{\bf {10}}^2  \delta^i_j
                             - R_{\bf {10}} V^{\dagger}_{j3} V^{3i}.
\end{eqnarray}

As pointed out in Ref.~\cite{BH} the radiative corrections from
the top quark Yukawa couplings produce large LFV masses for right-handed 
sleptons ($\tilde{l}_R$). On the other hand, the LFV masses for left-handed 
sleptons ($\tilde{l}_L$) induced by the neutrino Yukawa couplings 
\cite{MB,HMTYY,HMTY} seem to be relatively smaller 
than those for $\tilde{l}_R$ even if the right-handed tau-neutrino Yukawa 
coupling is comparable to the top quark Yukawa coupling since they are 
generated by color-singlet loops.\footnote
{The running effect between the GUT scale and the right-handed neutrino
mass makes the LFV masses for left-handed slepton larger. We include this
effect in our numerical calculations.}
However, the $(3,2)$ element of $U$ suggested by the atmospheric neutrino
anomaly is much larger than that of $V\sim 0.04$, and thus the left-handed 
sleptons can have larger LFV masses. 

In the above discussion, we ignore the bottom quark Yukawa coupling assuming
that the vacuum angle 
$\tan\beta(\equiv\langle H_2\rangle/\langle H_1\rangle)$ is not large. If
it is not the case, the LFV masses for both the left-handed and right-handed 
sleptons are slightly reduced. However, in this case the LFV event rates 
become larger as seen later since they are  proportional to  $\tan^2\beta$ 
when $\tan\beta\gsim 1$ \cite{HMTYY,HMTY,HMTYSU5}. 

Let us now discuss rates of LFV processes. First, we consider 
$\mu^+ \rightarrow {\rm e}^+ \gamma$ decay. The amplitude takes a form
\begin{eqnarray}
T=e \epsilon^{\alpha*}(q) \bar{v}_{\mu} (p) 
i \sigma_{\alpha \beta} q^\beta (A_L P_L + A_R P_R)
v_{\rm e}(p-q),
\label{Penguin}
\end{eqnarray}
where $p$ and $q$ are momenta of muon and photon.
Then, the decay rate is given by
\begin{eqnarray}
\Gamma(\mu \rightarrow {\rm e}\gamma)
= \frac{e^2}{16 \pi} m_{\mu}^3 (|A_L|^2+|A_R|^2).
\label{eventrate}
\end{eqnarray}
Since the operators in Eq.~(\ref{Penguin}) break not only the $SU(2)\times 
U(1)$ gauge symmetry but also a chiral symmetry of lepton,
$A_L$ and $A_R$ have to be proportional to a Yukawa coupling constant for
lepton and one of two vacuum expectation values of Higgs bosons.
Then,  the diagrams proportional to $\langle H_2 \rangle$ are enhanced
by $\tan\beta$ compared with those proportional to $\langle H_1 \rangle$,
as discussed in Refs.~\cite{HMTY,HMTYSU5}.
Furthermore, when both the left-handed and right-handed sleptons have
LFV masses, there are diagrams  proportional to the tau-lepton 
mass $m_{\tau}$ (see Fig.~1(a) and (b)).
This is a crucial point in the present model. In the SUSY $SU(5)_{GUT}$ model 
without the right-handed neutrinos, $A_R$ is proportional to $m_\mu$,
and $A_L$ is negligible since only the right-handed sleptons 
have LFV masses \cite{BH,HMTYSU5}. On the other hand, in the SUSY standard 
model
with the right-handed neutrinos, $A_L$ is proportional to $m_{\mu}$ and
$A_R$ negligible \cite{HMTYY,HMTY}. This is one of the reasons why we obtain 
relatively large rates for LFV decay processes. A similar enhancement
also appears in the SUSY $SO(10)_{GUT}$ model \cite{so10}. However,
the large mixing angle in the lepton sector required from the 
atmospheric neutrino anomaly is hard to be explained in the $SO(10)_{GUT}$
model without a fine tuning.

A dominant contribution to $A_R$ comes from the diagrams in Fig.~1(a), 
which are given by
\begin{eqnarray}
A_R &\simeq& -\frac{g^2_Y}{16\pi^2} m_{\tau} M_{\tilde{B}}\mu_H \tan\beta
(m_{\tilde{l}_L}^2)^2_3 (m_{\tilde{l}_R}^2)^3_1 
\nonumber\\
&&
D^3\left[\frac{1}{m^2}f(M_{\tilde{B}}^2/m^2);m^2\right]
(m_{\tilde{\mu}_L}^2,m_{\tilde{\tau}_L}^2,
 m_{\tilde{\tau}_R}^2,m_{\tilde{{\e}}_R}^2).
\label{aral}
\end{eqnarray}
where $\mu_H$ and $M_{\tilde{B}}$ are Higgsino and bino masses, and
$m_{\tilde{\mu}_L}$, $m_{\tilde{\tau}_L}$, $m_{\tilde{\tau}_R}$, and 
$m_{\tilde{{\e}}_R}$ are left-handed smuon and stau masses and 
right-handed stau and selectron masses.
We have assumed $\tan\beta \gsim 1$ and the SUSY breaking scale is 
larger than $Z$-boson mass $m_Z$ in derivation of Eq.~(\ref{aral}).
Here, a finite-difference function $D[f(x);x]$ is defined as
\begin{equation}
D[f(x);x] (x_0,x_1)
\equiv
\frac{1}{x_0-x_1}\left[f(x_0)-f(x_1)\right],
\end{equation}
and then, the $N$-th finite-difference function is given by
\begin{eqnarray}
D^{N}[f(x);x](x_0,x_1,\cdots,x_N)  
&=&
\sum^N_{i=0} \left(\prod_{j \ne i} \frac{1}{x_i -x_j}\right) f(x_i).
\end{eqnarray}
The function $f(x)$ is defined as
\begin{eqnarray}
f(x) &=& -\frac1{2(1-x)^3} (1-x^2 +2 x\log(x)).
\end{eqnarray}
We see that the dominant contribution to $A_R$ is proportional to
$m_{\tau}$. In estimating $A_R$ we use the maximal mixing $U^{32} 
\simeq 1/\sqrt{2}$
as suggested from the atmospheric neutrino anomaly.\footnote{
To calculate $A_R$ we also adopt the small angle MSW solution \cite{MSW}
to the solar neutrino problem, which gives 
$ |U^{32}| \simeq |U^{23}| \simeq |U^{22}| \simeq |U^{33}| \simeq 
\frac{1}{\sqrt{2}} $ \cite{Yasuda}.}
Due to this contribution the decay rate is enhanced by 
$(m_\tau/m_\mu)^2(U^{32}/V^{32})^2\sim 10^5$ compared with 
that in the SUSY $SU(5)_{GUT}$ model without the right-handed 
neutrinos \cite{BH,HMTYSU5}, if the Yukawa coupling for tau neutrino is 
as large as that of top quark ({\it i.e.} $f_{\nu_3}=f_{u_3}$).

There is a contribution to $A_L$ proportional to $m_\tau$,
which comes from Fig.~1(b), and the explicit form is given by
exchanging $L$ and $R$ in Eq.~(\ref{aral}). However, since the matrix 
element $U^{ 31}$ is unknown, 
$A_L$ has large ambiguity.\footnote{
If $(g_2^2/g_Y^2)  (m_{\mu}/m_{\tau})  (M_{\tilde{W}}/M_{\tilde{B}})
(U^{31}/V^{31}) \gg 1$ where $M_{\tilde{W}}$ is the wino mass, the wino 
contribution may be dominant in $A_L$.
}
Therefore, we take $A_L=0$ to  evaluate the lower bound of the decay
rate of  $\mu^+ \rightarrow {\rm e}^+\gamma$ from Eq.~(\ref{eventrate}).

Fig.~2 shows the dependence of the lower bound of $\Br(\mu\rightarrow 
\e \gamma)$ on the right-handed neutrino mass $M_N$.
In this and the following numerical calculations we adopt the requirement 
described in Ref. \cite{HMTY} which determines $\mu_H$ in terms of other 
parameters.  In this figure $m_{\tilde{\e}_R}=150$GeV, $M_{\tilde{B}}=50$GeV, 
and $\mu_H>0$.\footnote{Our result is insensitive to the sign of $\mu_H$ as  
seen from Eq.~(\ref{aral}).
}
Also, we have taken the top quark mass $m_{t}=175$GeV
and $m_{\nu_\tau}=0.1$eV (solid lines), and 
$m_{\nu_\tau}=0.03$eV 
(dashed lines). We take $a_0=0$ in Eq.~(\ref{InitCondAtGravScale}) for 
simplicity hereafter.   The larger decay rates
correspond to the larger $\tan\beta$ as we explained above. 
To demonstrate this we have taken $\tan \beta = 30,10$, and 3. 
Note that lower bounds for $\tan\beta=3$ are enhanced by relatively larger 
$\mu_H$
which is determined by the radiative breaking condition for the 
electroweak symmetry.

For a given tau-neutrino mass, the decay rate is almost proportional to
a square of $M_N$. It may be surprising that the prediction already exceeds
the experimental bound ($\Br(\mu\rightarrow \e \gamma)< 4.9\times 10^{-11}$
 \cite{PDG}) in a broad parameter region. We have restricted our analysis 
to the case where the perturbative description of the 
SUSY GUT is valid up to the gravitational scale. This is the reason why 
the prediction lines have cuts in the large $M_N$ region, $M_N \gsim 5\times 
10^{14}$GeV ($2\times 10^{15}$GeV) for $m_{\nu_\tau}= 0.1$
eV (0.03eV).  

Fig.~3 shows the dependence of the lower bound of $\Br(\mu\rightarrow 
\e \gamma)$ on $m_{\tilde{\e}_R}$ when the tau-neutrino Yukawa coupling 
is as large as that of the top quark at the gravitational scale ({\it i.e.}
$f_{\nu_3}=f_{u_3}$). We have fixed $m_{\nu_\tau}= 0.07$eV
and taken $M_{\tilde{B}}=50$GeV (solid lines) and $100$GeV(dashed lines).
In this figure the lower bounds for $\tan\beta=3$ are almost degenerate 
with those for
$\tan\beta=10$, and they are slightly larger in larger $m_{\tilde{\e}_R}$.
This comes from the fact that $\mu_H$ and $M_N$ for 
$\tan\beta=3$ are larger than those for $\tan\beta=10$. Also, in a region where
$m_{\tilde{\e}_R}$ is comparable to $M_{\tilde{B}}$,
the decay rate is suppressed since  the LFV masses are reduced by the 
radiative corrections from the gaugino loops. On the other hand, 
if $m_{\tilde{\e}_R}\gg M_{\tilde{B}}$, the decay rates are proportional to
square  of $M_{\tilde{B}}$ as seen in Eq.~(\ref{aral}). From this figure,
we see that $\Br(\mu\rightarrow \e \gamma)$ is larger than $\sim 10^{-14}$
when the tau-neutrino Yukawa coupling is as large 
as that for top quark. If the Yukawa coupling for tau neutrino is 
three times smaller than that of top quark, the lower bound is reduced
by a factor $\sim 100$.\footnote{
This smaller value of the Yukawa coupling for tau neutrino also reduces the 
Majorana mass for the right-handed neutrino by a factor $\sim 10$ for a given
$m_{\nu_{\tau}}$.} 

Next, we discuss other LFV processes for muon, $\mu^+\rightarrow \e^+\e^-\e^+$ 
and $\mu-\e$ conversion on $^{48}_{22}{\rm Ti}$. For these 
processes the photon penguin contributions in Figs.~1(c) and (d), where
large blobs denote  the operators in Eq.~(\ref{Penguin}), give dominant 
amplitudes. Then, we can also derive the lower bounds of 
$\Br(\mu\rightarrow 3\e)$ and 
${\rm R}(\mu^-\rightarrow\e^-;~^{48}_{22}{\rm Ti}) $\footnote{
The $\mu-\e$ conversion rate ${\rm R}(\mu^-\rightarrow\e^-;~^{48}_{22}
{\rm Ti})$ is normalized by the muon capture rate. For the detailed definition
of R, see Ref.~\cite{PDG}.}
from $\Br(\mu\rightarrow \e \gamma)$, which are given by
\begin{eqnarray}
\Br(\mu\rightarrow 3\e)&\simeq& 7 \times 10^{-3}  \Br(\mu\rightarrow \e \gamma),
\nonumber\\ 
{\rm R}(\mu^-\rightarrow \e^-;~^{48}_{22}{\rm Ti})&\simeq& 6 \times 10^{-3} \Br(\mu\rightarrow \e \gamma),
\end{eqnarray}
Thus, we find that $\Br(\mu\rightarrow 3\e)$ and 
${\rm R}(\mu^-\rightarrow\e^-;~^{48}_{22}{\rm Ti}) $ are larger than
$\sim 10^{-16}$.  

Finally, we discuss $\tau^\pm \rightarrow \mu^\pm \gamma$ decay. For this
process the amplitude is given by  the same form as Eq.~(\ref{Penguin}),
and $A_L$ and $A_R$ depend on $(m_{\tilde{l}_L}^2)^3_2$ 
and $(m_{\tilde{l}_R}^2)^3_2$, which are both calculable since 
$U^{32}$
and $V^{32}$ are known. We assume again that the tau-neutrino 
Yukawa coupling is as large as that of the top quark at the gravitational
scale $M_G$. In this case we find that the amplitude $A_L$ is larger than $A_R$
in most of the parameter space. This comes from the fact that 
$(m_{\tilde{l}_L}^2)^2_3$ is larger than $(m_{\tilde{l}_R}^2)^2_3$. 
When the SUSY breaking scale is larger than $m_Z$, $A_L$ is given by
\begin{eqnarray}
A_L &\simeq& \frac{g_2^2}{16 \pi^2} m_{\tau} M_{\tilde{W}}\mu_H \tan\beta  
        (m_{\tilde{l}_L}^2)^2_3
\nonumber\\
&&
        D\left[
        D\left[\frac{1}{m^2}g(M^2/m^2);M^2\right](M_{\tilde{W}}^2,\mu^2)
           ;m^2
         \right]( m_{\tilde{\mu}_L}^2,m_{\tilde{\tau}_L}^2),
\end{eqnarray}
where $M_{\tilde{W}}$ is the wino mass and we have assumed $\tan\beta\gsim 1$.
The function $g(x)$ is 
\begin{eqnarray}
g(x) &=& -\frac1{4(1-x)^3}(7-8x+x^2 +2(2+x)\log(x)).
\end{eqnarray}
Fig.~4 shows the dependence of $\Br(\tau \rightarrow \mu \gamma)$ on 
$m_{\tilde{\e}_R}$. The parameters are taken as in Fig.~3. We see that the
 branching ratio of $\tau^{\pm} \rightarrow \mu^{\pm} \gamma$ is larger 
than $10^{-9}$ for a broad parameter region.

In conclusion, we have discussed the LFV processes at low energies in the SUSY
 $SU(5)_{GUT}$ model with the right-handed neutrinos.
The atmospheric neutrino anomaly reported by the super-Kamiokande collaboration
suggests a large LFV Yukawa coupling at the GUT scale which induces naturally
large LFV at low energies through radiative corrections. We have, in fact, 
found that lower bounds for $\Br(\mu \rightarrow e \gamma)$, 
$\Br(\tau \rightarrow \mu \gamma)$, and 
${\rm R}(\mu^-\rightarrow\e^-;~^{48}_{22}{\rm Ti})$
reach to $10^{-14}$, $10^{-9}$, and $10^{-16}$,
respectively, for a wide range of parameter space, if the Yukawa coupling for 
tau neutrino is as large as that of top quark. This assumption ($f_{\nu_3}=
f_{u_3}$) corresponds to $M_N \simeq (1-2) \times 10^{14}$GeV for 
$m_{\nu_{\tau}}=0.07$eV. This Majorana mass for right-handed neutrino
is quite naturally obtained from Eq.~(\ref{m_rn_mass}) with $c\simeq O(1)$.

 In the future experiments lower bounds for
$\Br(\mu \rightarrow e \gamma)$ and 
${\rm R}(\mu^-\rightarrow\e^-;~^{48}_{22}{\rm Ti})$ are expected to come
down to $10^{-14}$ \cite{kuno,mega,Orito} and $10^{-16}$ \cite{meco}, 
respectively. We hope that the LFV processes also provide us with 
strong indication of the presence of right-handed neutrinos. 

\underline{Acknowledgment}

One of the authors (J.H.) would like to thank Y.~Okada and Y.~Kuno.

\newpage
%%%%%%%%%%%%%%%%%%%%%%%%%%%%%%%%%%%%%%%%%%%%%%%%%%%%%%%%%%%%%%%
%
% NEW COMMANDS FOR THE BIBLIOGRAPHY
%
%%%%%%%%%%%%%%%%%%%%%%%%%%%%%%%%%%%%%%%%%%%%%%%%%%%%%%%%%%%%%%%
\newcommand{\Journal}[4]{{\sl #1} {\bf #2} {(#3)} {#4}}
\newcommand{\APJ}{Ap. J.}
\newcommand{\CJP}{Can. J. Phys.}
\newcommand{\NC}{Nuovo Cimento}
\newcommand{\NP}{Nucl. Phys.}
\newcommand{\PL}{Phys. Lett.}
\newcommand{\PR}{Phys. Rev.}
\newcommand{\PRep}{Phys. Rep.}
\newcommand{\PRL}{Phys. Rev. Lett.}
\newcommand{\PTP}{Prog. Theor. Phys.}
\newcommand{\SJNP}{Sov. J. Nucl. Phys.}
\newcommand{\ZP}{Z. Phys.}
%%%%%%%%%%%%%%%%%%%%%%%%%%%%%%%%%%%%%%%%%%%%%%%%%%%%%%%%%%%%%%%

\newpage
%
%
%%%%%%%%%%%% Fig.1%%%%%%%%%%%%%%%%%%%%%%
%
%
\begin{figure}[p]
\epsfxsize=15cm
%\centerline{}
\caption{
(a) A contribution to $A_R$, proportional to 
$m_\tau \tan\beta$.
The large blobs are lepton-flavour violating masses of sleptons.
(b) A contribution to $A_L$, proportional to 
$m_\tau \tan\beta$.
The large blobs are lepton-flavour violating masses of sleptons.
(c) A contribution to $\mu^+\rightarrow\e^+\e^-\e^+$ from lepton-flavour 
violating dipole terms. The large blob here denotes lepton-flavour 
violating dipole terms.
(d) A contribution to $\mu-\e$ conversion from lepton-flavour 
violating dipole terms. The large blob denotes lepton-flavour 
violating dipole terms. }

\end{figure}
%
%
%%%%%%%%%%%% Fig.2%%%%%%%%%%%%%%%%%%%%%%
%
%
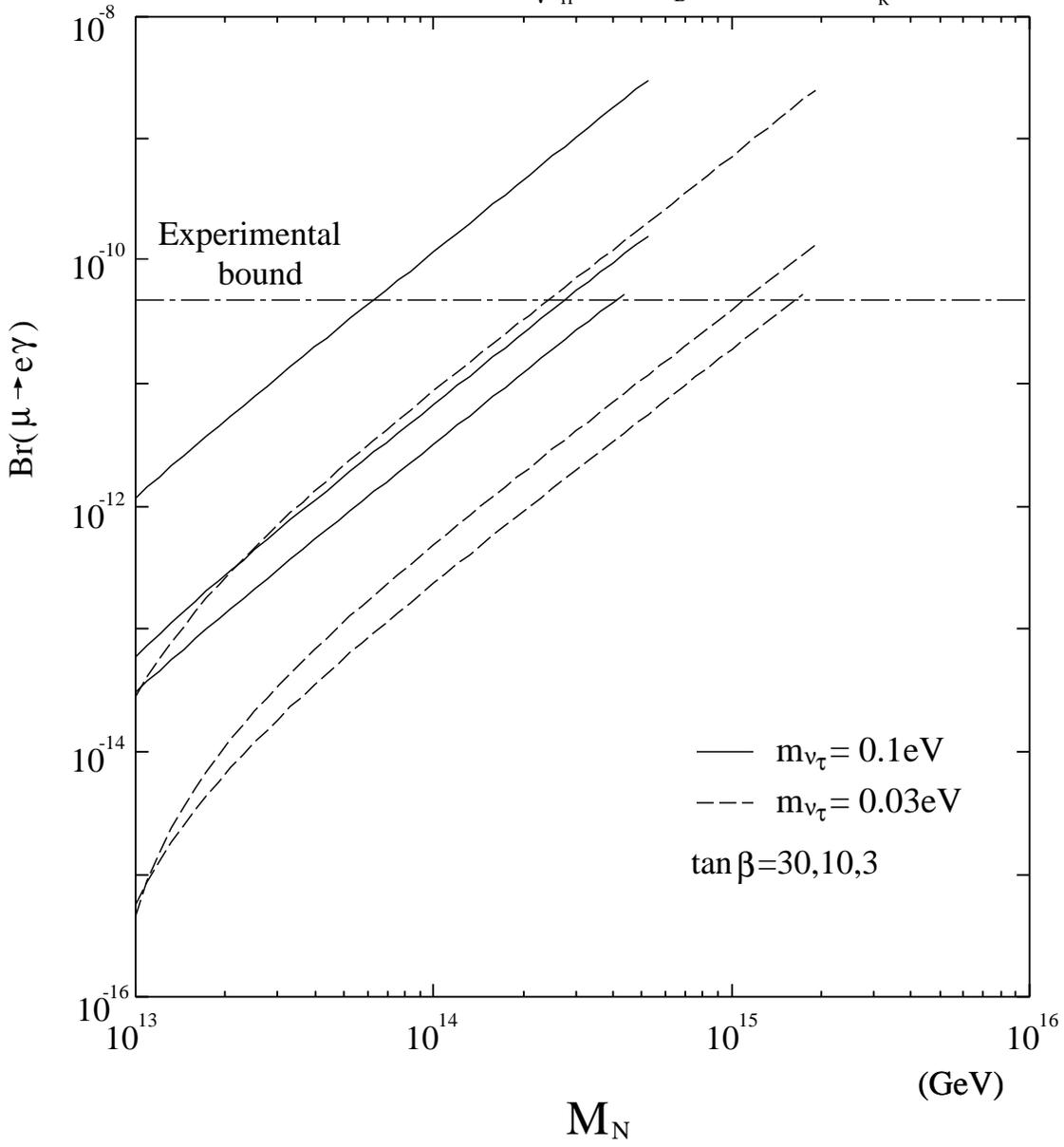
\begin{figure}[p]
\epsfxsize=15cm
%\centerline{}
\caption
{
Dependence of the lower bounds of $\Br(\mu\rightarrow  \e \gamma)$ on the 
right-handed neutrino mass $M_N$. In this figure the right-handed selectron 
mass $m_{\tilde{\e}_R}=150$GeV, the bino mass $M_{\tilde{B}}=50$GeV, and  
the Higgsino mass $\mu_H>0$. The solid lines are for $m_{\nu_\tau}=0.1$eV, 
and the dashed lines for $m_{\nu_\tau}=0.03$eV.
We take $\tan \beta = 30,10$, and 3, and the larger decay rates correspond 
to the larger $\tan\beta$. 
}
\end{figure}
%
%
%%%%%%%%%%%% Fig.3%%%%%%%%%%%%%%%%%%%%%%
%
%
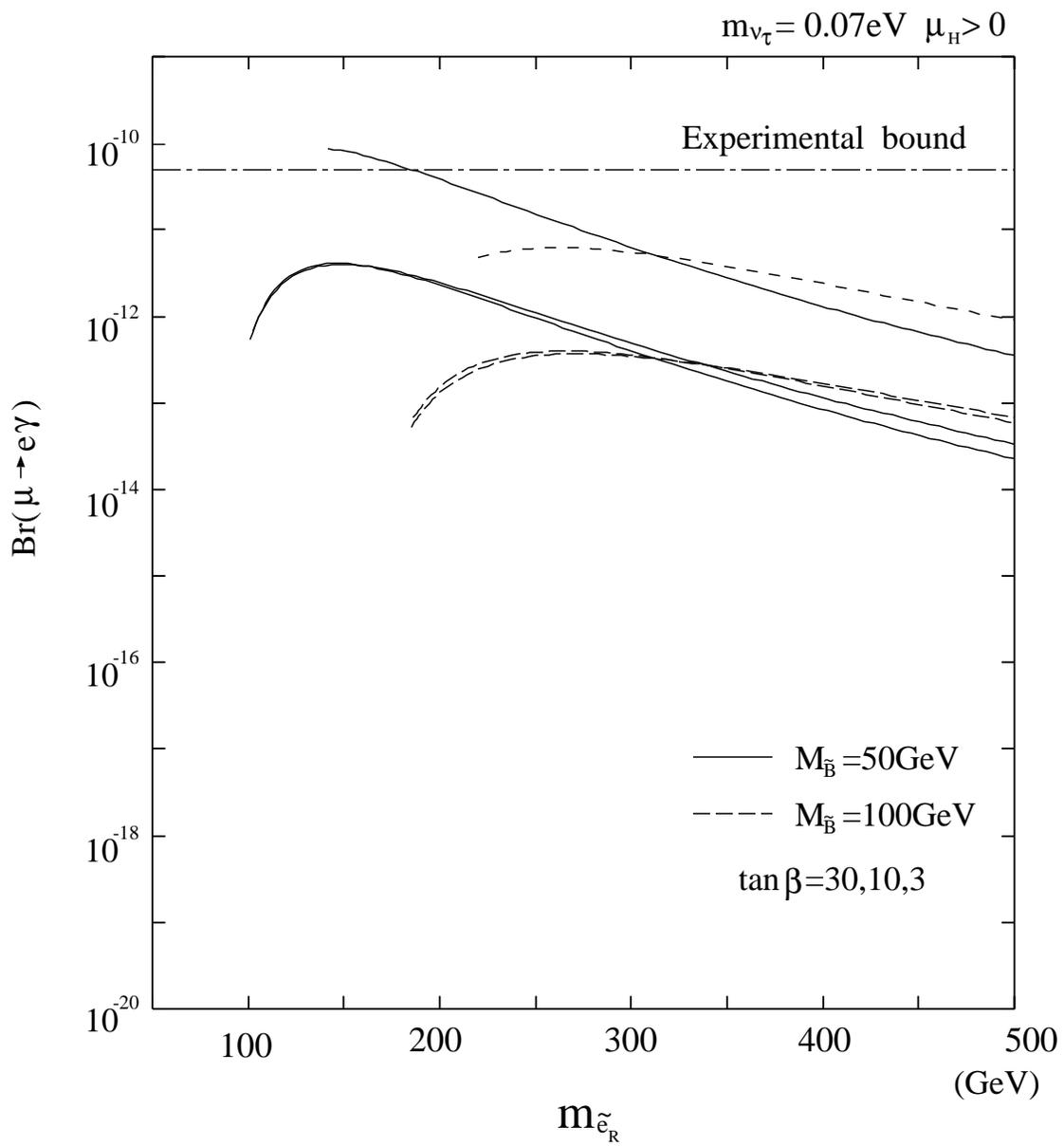
\begin{figure}[p]
\epsfxsize=15cm
%\centerline{}
\caption
{
Dependence of the lower bounds of $\Br(\mu\rightarrow 
\e \gamma)$ on the right-handed selectron mass  $m_{\tilde{\e}_R}$ when the 
tau-neutrino Yukawa coupling is as large as that of the top quark at the 
gravitational scale $\simeq$ ($2\times 10^{18}$GeV). We fix $m_{\nu_\tau}= 
0.07$eV. Here, the solid lines are for $M_{\tilde{B}}=50$GeV and  the
dashed lines are for $M_{\tilde{B}}=100$GeV.  We take $\tan \beta = 30,10$, 
and 3, and the lines for $\tan\beta=3$ are almost degenerate with those for
$\tan\beta=10$, and they are slightly larger in larger $m_{\tilde{\e}_R}$.
}
\end{figure}
%
%
%%%%%%%%%%%% Fig.4%%%%%%%%%%%%%%%%%%%%%%
%
%
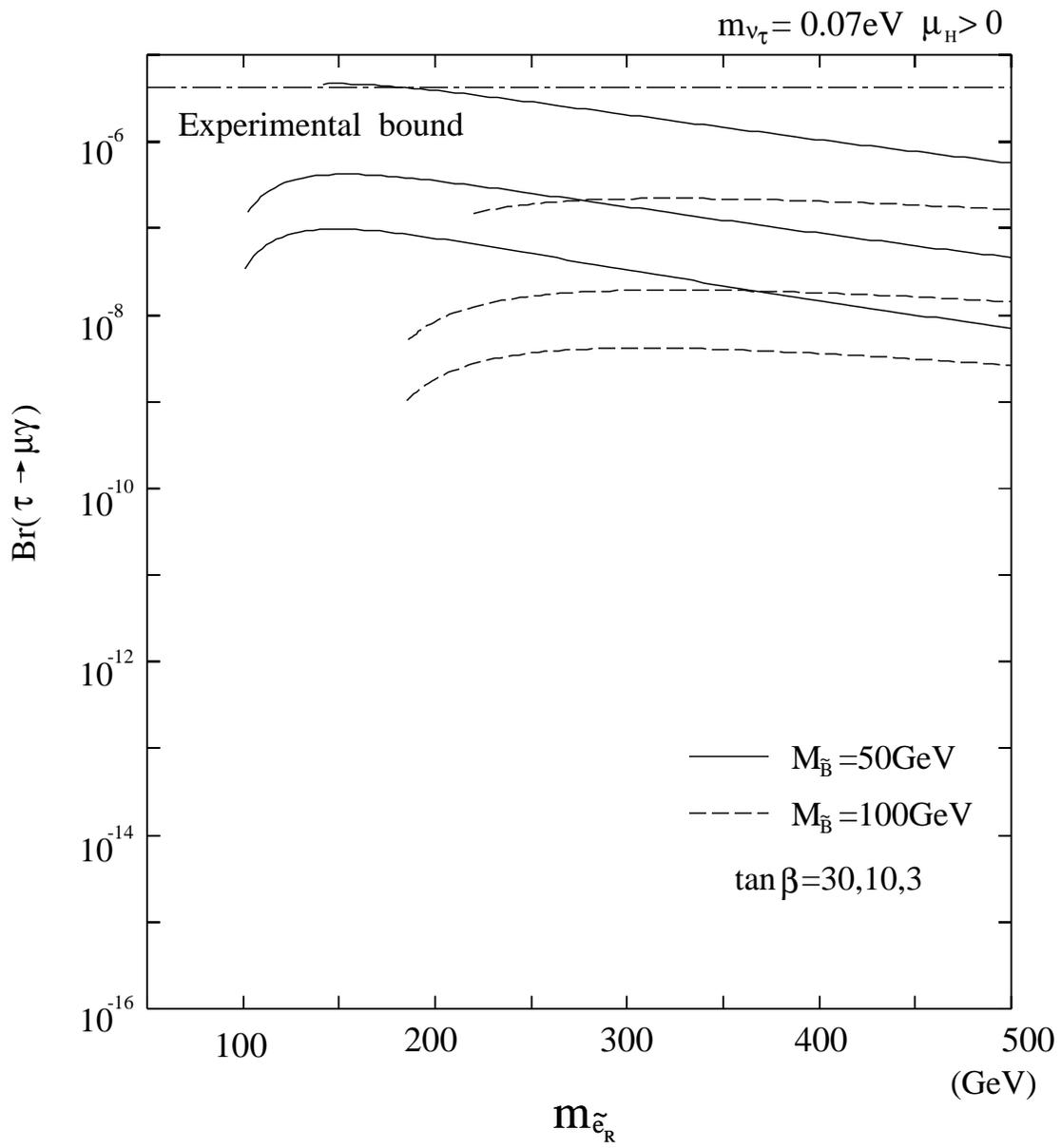
\begin{figure}[p]
\epsfxsize=15cm
%\centerline{}
\caption
{
Dependence of $\Br(\tau\rightarrow 
\mu \gamma)$ on the right-handed selectron mass  $m_{\tilde{\e}_R}$ when the 
tau-neutrino Yukawa coupling is as large as that of the top quark at the 
gravitational scale $\simeq$ ($2\times 10^{18}$GeV). We fix $m_{\nu_\tau}= 
0.07$eV. Here, the solid lines are for $M_{\tilde{B}}=50$GeV and  the
dashed lines are for $M_{\tilde{B}}=100$GeV.  We take $\tan \beta = 30,10$, 
and 3, and the larger decay rates correspond to the larger $\tan\beta$.
}
\end{figure}

\end{document}